\documentclass[english,aps,twocolumn,prX,showpacs,superscriptaddress]{revtex4-1}

\usepackage{graphicx}
\usepackage{dcolumn}
\usepackage{bm}
\usepackage{babel}
\usepackage[T1]{fontenc}
\usepackage[latin9]{inputenc}
\usepackage{amsmath}
\usepackage{amssymb}
\usepackage{amsfonts}
\usepackage{hyperref}
\setcitestyle{numbers,square}
\usepackage{color}
\hypersetup{
	colorlinks = true,
	pdftitle = {},
	pdfauthor = {},
	pdfkeywords = {},
	linkcolor = blue,
	citecolor = blue,
	filecolor = black,
	urlcolor = magenta,
}

\begin{document}
\title{Optical excitations and thermoelectric properties of 2D holey graphene}

\author{Deobrat Singh}
\email{deobrat.singh@physics.uu.se}
\affiliation{Condensed Matter Theory Group, Department of Physics and Astronomy, Uppsala University, Box 516, SE-75120, Uppsala, Sweden}
\author{Vivekanand Shukla}
\altaffiliation{Present Address: Department of Microtechnology and Nanoscience (MC2), Chalmers University of Technology, SE-41296 Gothenburg, Sweden}
\email{vns391@gmail.com}
\affiliation{Condensed Matter Theory Group, Department of Physics and Astronomy, Uppsala University, Box 516, SE-75120, Uppsala, Sweden}
\author{Rajeev Ahuja }
\affiliation{Condensed Matter Theory Group, Department of Physics and Astronomy, Uppsala University, Box 516, SE-75120, Uppsala, Sweden}
\affiliation{Applied Materials, Department of Materials and Engineering, Royal Institute of Technology (KTH), SE-10044 Stockholm, Sweden}

\date{\today}


\begin{abstract}
Recently, holey graphene (HG) has successfully synthesized at atomic precission of hole size and shape. This shows interesting physical and chemical properties for energy and environmental applications. Shaping of the pores also transforms semimetallic graphene to semiconductor holey graphene, which opens new door for its use in electronic applications. We systematically investigated the structural, electronic, optical and thermoelectric properties of HG structure using first-principles calculations. HG was found to have a direct band gap with 0.65 eV (PBE functional), 0.95 eV (HSE06 functional) and HSE06 functional is in good agreement with experimental results. For the optical properties, we use single-shot G0W0 calculations by solving the Bethe-Salpeter equation to determining the intralayer excitonic effects. From the absorption spectrum, we obtained the optical gap of 1.28 eV and a week excitonic binding energy of 80 meV. We have found the large values of thermopower of 1662.59 $\mu$V/K and  better electronic figure of merit, ZT$_{e}$ as 1.13 from the investigated thermoelectric properties. Our investigations exhibit strong and broad optical absorption in the visible light region, which makes HG monolayer a promising candidate for optoelectronic and thermoelectric applications.
\end{abstract}

\maketitle


\section{Introduction}\label{sec:intro}

The first acknowledged isolation and discovery of single-layer graphene by backward exfoliation of graphite was accomplished in 2004 by Geim and coworkers\cite{Gr-1,Gr-2}. Since then, significant efforts to study the captivating attributes, applications, and synthesis methods of graphene have commenced to new research directions\cite{Gr-3,Gr-4}. Graphene consists of a monolayer of carbon atoms in a honeycomb arrangement with sp$^{2}$ hybridized carbon atoms that exhibit unusual and interesting electrical, optical, thermal, and mechanical characteristics. Along with other properties, graphene has resulted in increased research attention for realizing its applicability on a realistic scale\cite{Gr-5,Gr-6,Gr-9}. It is well understood that graphene manifests high electrical conductivity, a frictionless surface, a high ambipolar electric field and quantum hall effects at room temperature. Besides, graphene is believed to be the strongest material ever known to humankind and exhibits high carrier mobility\cite{Gr-7,Gr-8,Gr-10}. Graphene is an excellent candidate for transparent and conductive composites and electrodes, thin-film transistors, and photonics applications due to these unique properties. However, despite having all these unique properties, the intrinsic dispersion of graphene possesses a zero-gap electronic structure. This leads to limitations for its use in some electronic applications such as transistors, where it exhibits extremely high off-state current and very low on/off ratio\cite{Gr-11,Gr-12}. 

This particular limitation of graphene acted positively and served as motivation for scientific endeavors to synthesize, design, and fabricate novel 2D materials during the last one decade\cite{2D-1,2D-2,2D-3}. We got full family or 2D materials with a variety of different properties and applications. Apart from this, several physical and chemical strategies such as oxidation, hydrogenation and fluorination have been tried out to produce an electronic gap in graphene pertaining to its high mobility\cite{2D-4,2D-5,2D-6,2D-7}.

In recent years, the perforation of 2D materials emerged as effective maneuvering to enhance electronic, mechanical, optical properties and broaden the applicability exceeding its pristine structure\cite{Perfo-1,Perfo-2,Perfo-3,Perfo-4}. Perforation of graphene, which is also known as holey graphene (HG) with small holes turns out to be permeable, and this change, combined with graphene's intrinsic strength and layered thickness, directs to its future application as the most flexible, and selective filter for tiny substances including biomolecules, greenhouse gases, and salts\cite{Perfo-5,Perfo-6,Perfo-7}.  Exceeding this when the spacing between holes is repeatedly reduced to a few atoms transforms its electronic structure from semimetal to semiconductor\cite{Perfo-12}. There have been several theoretical reports in various HG with different pore sizes\cite{Perfo-8,Perfo-9,Perfo-10,Perfo-11}. HG can have astonishing properties owing to the quantum confinement effect, although the controlled experimental realization of HG with finite perforation remains a challenging task. Recently group in Spain has ({\it{Moreno et al. Science 2018,360, 199}}) has devised a way to synthesize the holey graphene with outstanding precision in bottom up on surface method starting from organic molecules\cite{moreno2018bottom}. Apart from this, they also have been successful in transferring it to the dielectric substrate, which enhances the possibility of using the HG for possible device applications where it can replace the bulkier, more rigid silicon components used today.
 
Nowadays, due to their unexceptionable electronic transport properties, the two-dimensional (2D) materials have demonstrated superior potential applications in thermoelectric (TE) energy generation, and direct conversion of heat to electricity. The properties of TE materials are characterized by the dimensionless figure of merit (ZT), ZT = $\sigma$ S$^{2}$T/($\kappa_{e}+\kappa_{l})$, where S is the Seebeck coefficient, $\sigma$ is electric conductivity, T is absolute temperature and $\kappa_{e}$ and $\kappa_{l})$  shows the electronic thermal conductivity and lattice thermal conductivity in which lattice thermal conductivity relative with lattice vibrations (phonon) and $\kappa_{e}$ is electronic thermal conductivity related with electronic structure. The large values of ZT for TE materials are important for practical thermoelectric device applications. Past few years the intensive research work has been done to enhanced the ZT values of the nanostructured manterials for high thermoelectric performance\cite{dresselhaus2007new, zeier2016engineering, heremans2013thermoelectrics, haskins2011control}.

Taking to this as a motivation, in the present work, we study HG structural electronic optical and thermoelectric  properties with hybrid functional HSE06. Further, we also investigated G0W0+BSE method for excitonic optical properties, since to our knowledge there is no work has been done on this HG with G0W0+BSE hence this work is timely. The GW+BSE approach is much more accurate for optical properties for excitonic effect due to its quasi particle treatment. In the previous investigations \cite{moreno2018bottom} suggest that the QP band gap of HG monolayer is overestimated as compare to experimental band gap. In this study, the HSE06 functional gives more accurate band gap and it is consistent with experimental work. We found that the strong optical adsorptions in the visible light region for HG monolayer. Also in thermoelectric properties, we have calculated the Seebeck coefficient (thermopower), electrical conductivity, electronic thermal conductivity and electronic figure of merit. Our results suggest that novel 2D planar HG has better visible light absorption which makes 2D HG a promising candidate for potential applications in the field of opto and thermo electronic devices. 

 This paper is categorized in the following sections. In the section \ref{method}, we describe the computational methods employed in the present work. In section \ref{stru-elec}, we study the structural and electronic properties of HG monolayer. Section \ref{optical} shows the excitonic effect, and investigates the optical proprieties of HG monolayer. Finally, based on electronic transport, section \ref{thermo} explores the thermoelectric properties of HG monolayer.


\section{Computational method}
\label{method}
The electronic structure are based on first-principles calculations and we employed the plane-wave basis projector augmented wave (PAW) method in the framework of density-functional theory (DFT)\cite{hohenberg1964inhomogeneous}. For the exchange-correlation potential, the generalized gradient approximation (GGA) in the  form of Perdew-Burke-Ernzerhof (PBE) [41,42] was employed as implemented in the Vienna Ab Initio Simulation Package (VASP) software\cite{kresse1996efficient}. The inherent underestimation of the band gap given by DFT within the inclusion of hybrid functional is corrected by using the Heyd-Scuseria-Ernzerhof (HSE06)\cite{heyd2003hybrid} screened-nonlocal-exchange functional of the generalized Kohn-Sham scheme. The charge transfers in the structures was determined by the Bader analysis\cite{henkelman2006fast}. The energy cutoff value for the plane-wave basis set was taken to be 500 eV. The total energy was minimized until the energy variation in successive steps became less than 10$^{-6}$ eV in the structural relaxation and the convergence criterion for the Hellmann-Feynman forces was taken to be 10$^{-3}$ eV/$\AA$. 2x4x1 $\Gamma$ centered k-point sampling is used for the primitive unit cell. The Gaussian broadening for the density of states calculation was taken to be 0.10.

The dielectric function and the optical oscillator strength of the HG monolayer were calculated by solving the Bethe-Salpeter equation (BSE) on top of single-shot G0W0 calculation, which was performed over standard DFT calculations\cite{shishkin2006implementation, karlicky2013band}. The G0W0+BSE approach accounting for both e-e and e-h effects. Here, e-e and e-h represent the electron-electron and electron-hole correlation effects, respectively. During this process we used 2x4x1 $\Gamma$ centered k-point sampling. The cutoff for the response function was set to 250 eV. The number of bands used in our calculations is 340. The cutoff energy for the plane waves was chosen to be 400 eV. We included 36 valence (occupied orbitals) and 72 conduction (unoccupied orbitals) bands into the calculations for the dielectric function of HG monolayers in the BSE calculations. In previous investigations, 96 converged empty states for MoS2 monolayer\cite{qiu2013optical} and 150 empty states were taken for GW calculations for bulk system with semiconductors and insulators by Shishkin and co-workers\cite{shishkin2007self}.

Further the electron transport properties are computed by using semi-classical Boltzmann transport theory with the relaxation time approximation and rigid band approximation as implemented in the BoltzTraP code\cite{madsen2006boltztrap}. In the thermoelectric properties, we obtain the Seebeck coefficient (thermopower) $\alpha$ which is independent of the relaxation time $\tau$, electrical conductivity $\sigma$, electronic thermal conductivity $\kappa_{e}$ which depends on the relaxation time $\tau$. In order to evaluate the thermoelectric properties, we use the electronic figure of merit, ZT$_{e}$ = $S^{2}{\sigma T}$/$\kappa_{e}$, which shows the characteristics of electron transport and gives the upper limit of total ZT.
        

\section{Results and Discussion }
     \subsection{Structural and electronic properties} \label{stru-elec}
\begin{figure}[t]
	    \centering
	    \includegraphics[width=1.0\linewidth]{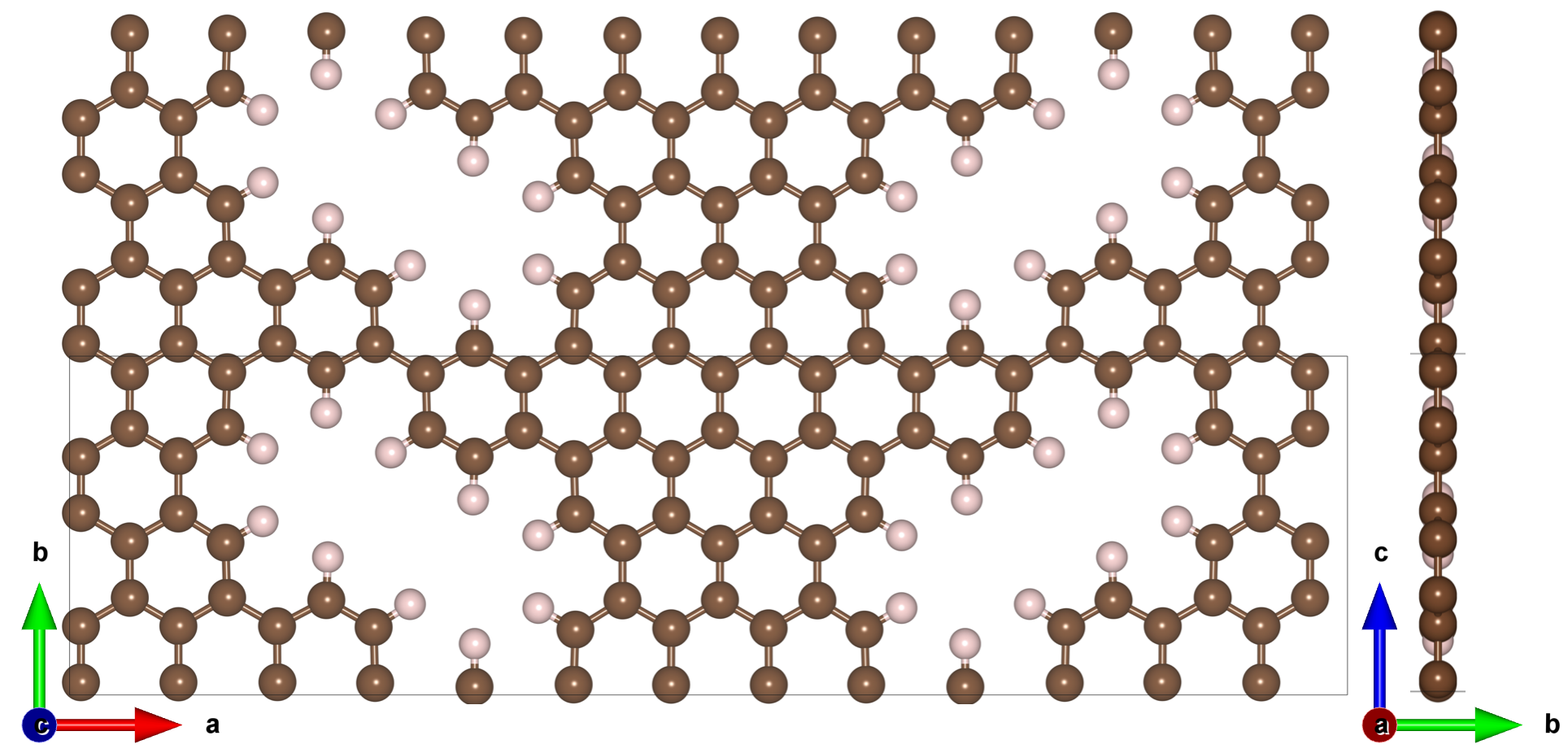}
	    \caption{Schematic representation of the optimized holey graphene with top and side view.}
	    \label{F1}
    \end{figure}

     The optimized structure of HG monolayer is depicted in Figure \ref{F1}. Due to the presence of two pore with different orientation, it has large unit cell with lattice parameters of a=32.383 \AA {} and b=8.583 \AA. Figure \ref{F1} shows the top and side view of HG structure which shows the planar structure. Also, we have calculated the bond length between C-C and C-H are 1.38 to 1.43 \AA {} and 1.09 \AA {} which is good consistent with previous work \cite{hussain2020functionalized, mortazavi2019nanoporous}. 
     
\begin{figure}[t]
	    \centering
	    \includegraphics[width=1.0\linewidth]{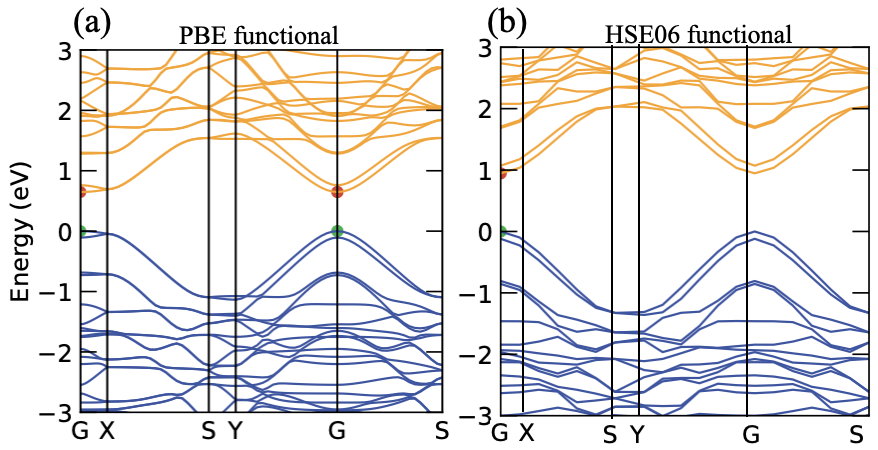}
	    \caption{Electronic band structures and corresponding projected density of states of holey graphene with (a) PBE functional and (b) HSE06 functional.}
	    \label{F2}
    \end{figure}

     Moving forward , the electronic band structure and corresponding projected density of states (PDOS) is presented in Figure \ref{F2}(a) using PBE functional. It is evident from the band structure that it has direct band gap of 0.65 eV at $\Gamma$ point. One can see that the p-orbital of C atom is mainly contributing near the Fermi level, E$_{F}$ in valence band maximum (VBM) and also in conduction band minimum (CBM) (see Figure \ref{F2}).There is no contribution from the hydrogen S-orbitals also appears in VBM and CBM region. The computed band gap is good consistent with previously investigated work \cite{moreno2018bottom, calogero2018electron, hussain2020functionalized, mortazavi2019nanoporous}. As per we know PBE generally underestimates the band gap of materials hence to find the more accurate electronic band structure, we use hybrid functional HSE06, which involves 25\% of Fock exchange (clearify this). The HSE06 functional give the band gap of 0.95 eV as shown in Figure \ref{F2}(b). which is well matched with the experimentally reported bandgap of this perforated monolayer.
     
\begin{figure}[t]
	    \centering
	    \includegraphics[width=1.0\linewidth]{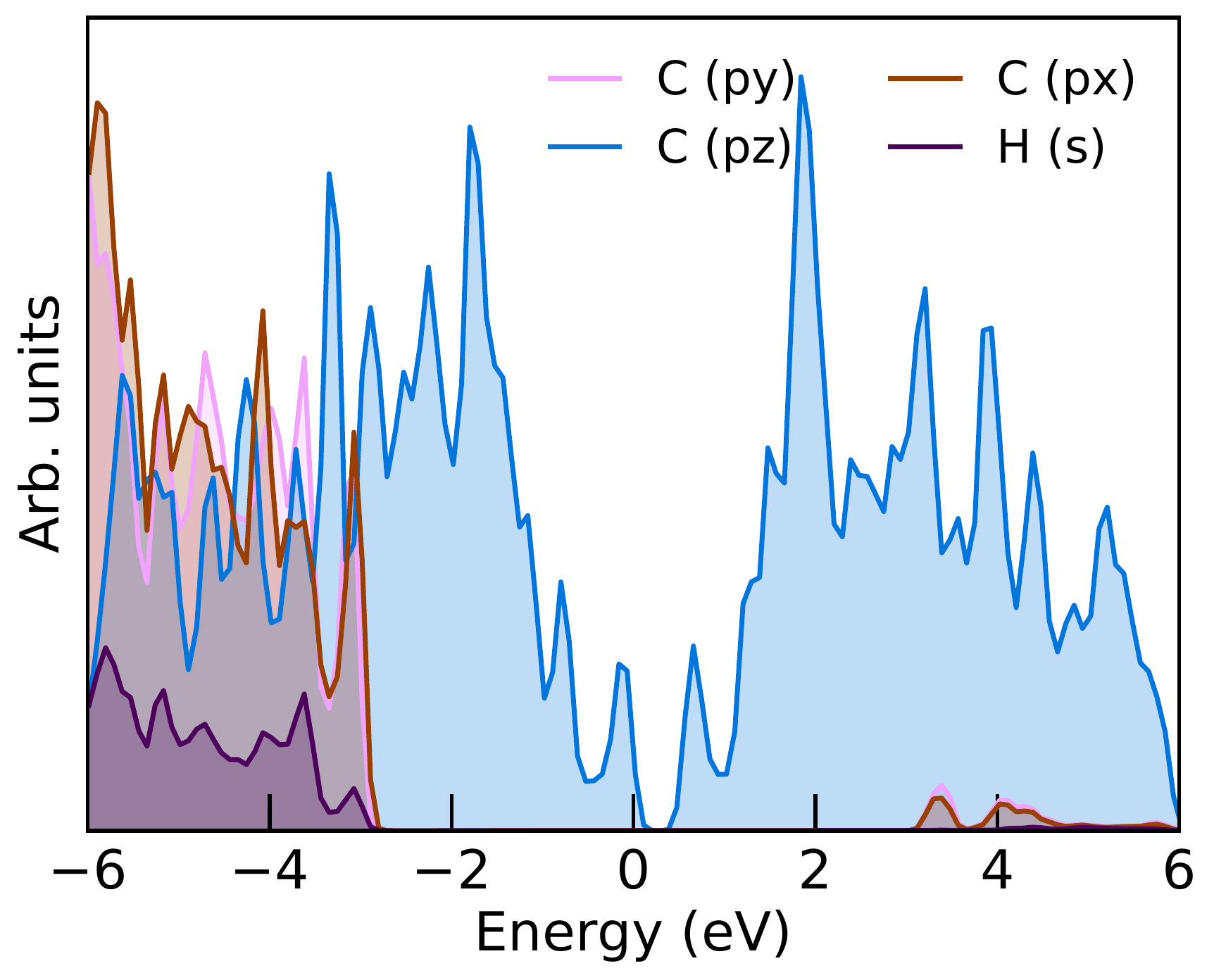}
	    \caption{Decomposed orbital density of states of HG monolayer with PBE functional functional.}
	    \label{orbital}
    \end{figure}

 The decomposed orbital density of states for HG monolayer are displayed in Figure \ref{orbital} showing three sub-bands: (1) the lower valence bands ($<$-3 eV) mainly dominated by C-px, py and H-s states; (2) the sub-bands dominated by C-pz states near (valence band) and above (conduction band) around the Fermi level; and (3) the strongly hybridized C-px, py with H-s sub-bands ($>3$ eV). These information is crucial and will help during the interband transitions involving from the valence band maximum (VBM) to conduction band minimum (CBM). The possible interband transition is mainly originates from (i) $\pi$ $\rightarrow$ $\pi^{*}$ with energy range of -3 eV to +3 eV, (ii) $\pi$ $\rightarrow$ $\sigma^{*}$, $\sigma$ $\rightarrow$ $\pi^{*}$ and $\sigma$ $\rightarrow$ $\sigma^{*}$ for other energy range. In the section \ref{optical}, we will take opportunity to  discuss about it.
 
Before moving to the possible optical transition, we have also calculated the effective mass and carrier mobility of electron and hole in HG monolayer. The simulated effective masses of the electrons ($m_{e}^{*} /m_{0}$) and holes ($m_{h}^{*} /m_{0}$) for the CBM and VBM of the HG monolayer is defined by;

\begin{equation}
m^{*}= \hbar^{2}\bigg[\frac{\partial^{2} E(k)}{\partial k^{2}}\bigg]^{-1}
\end{equation}

where, $\hbar$, k and E(k) represents the reduced plank constant, wave vector and the respective energy dispersion in CBM and VBM. Using the above equation the effective masses of electron and hole are estimated as 0.025 $m_{0}$ and 0.025 $m_{0}$ respectively. In previous report the effective mass of electron in graphene monolayer was found to be 0.012 $m_{0}$\cite{tiras2013effective}. Moreover, the theoretical carrier mobility for a better understanding of the electronic conductance of HG monolayer is calculated by using deformation potential (DP) theory devised by Bardeen and Shockley\cite{bardeen1950deformation}. The carrier mobility can be calculated by following relation\cite{mishra2020two};

\begin{equation}
\mu=\frac{2e\hbar^{3}C}{3k_{B}T\mid m_{*}\mid^{2}E_{1}^{2}},
\end{equation}
where, C is the elastic modulus which is calculated by quadratic fitting of the energy-strain data, k$_{B}$ and T are the Boltzmann constant and temperature, and $m^{*}$ is the effective mass calculated by using previous equation. $E_{1}$ is the deformation potential ( E$_{1}$ = $\Delta E$ ($\Delta a$/a), in which $\Delta E$ is the energy shift in the valence/conduction band edge with respect to the lattice variation ($\Delta a$/a). The calculated electron and hole carrier mobilities are 0.885x10$^{3}$ cm$^{2}$V$^{-1}$s$^{-1}$ and 10.8x10$^{3}$ cm$^{2}$V$^{-1}$s$^{-1}$, respectively. The carrier mobility of HG graphene is lower than the pristine graphene monolayer which is 3-4x10$^{5}$ cm$^{2}$V$^{-1}$s$^{-1}$ for electron and hole\cite{chen2013carrier, xi2012first, li2014intrinsic}. This is good for nanoelectronic device application because graphene shows poor on/off ratio due to its extremely high mobility and semi metallic nature.

    \subsection{Optical properties}
    \label{optical}
\begin{figure}[t]
	    \centering
	    \includegraphics[width=1.0\linewidth]{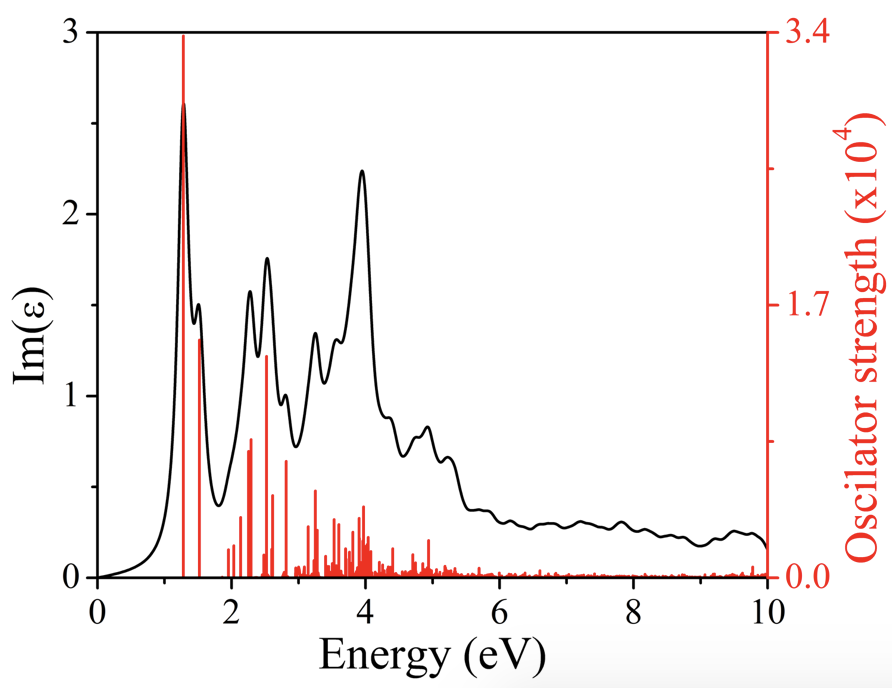}
	    \caption{Imaginary part of the dielectric function (black line) and the oscillator strength of the optical transitions (red bars) of holey graphene.}
	    \label{opt-transition}
    \end{figure}

In this section applying the GW+BSE method, the optical properties of HG monolayer, such as the real Re($\epsilon$) and imaginary part Im($\epsilon$) of the dielectric function, anisotropic nature of Im($\epsilon$), complex refractive index, absorption spectra, optical conductivity, electron energy-loss spectrum (EELS) and reflectively are computed.  The e-e correlation is considered to be the imaginary part Im($\epsilon$) of the dielectric function for HG by using G0W0 plus BSE functions. The G0W0 plus BSE, which establish higher order interaction illustration, i.e., e-e and e-h effects, is considered to get accurate the electronic description systematically on top of G0W0. Additionally, the e-h interaction produces mainly a re-normalization of the intensity of the optical peaks computed using the G0W0 plus BSE method. The imaginary part of the dielectric functions acquired at the G0W0 plus BSE levels for HG is shown in Figure \ref{opt-transition} , one can see that the inclusion of both e-e and e-h interactions yields a noteworthy red shift, which is good consistent with the previous results\cite{shahrokhi2016quasi}. Another appealing peculiar result is that the first BSE optical peak is in much better agreement with the electronic gap by GW method than the other two calculated results, implying the weakly bound excitonic/free carrier nature of the optical excitation. Consequently, the physical effect of the e-e and e-h interactions, reproduced by the G0W0 plus BSE method, provides a more accurate result. 

In order to examine the optical properties of the HG monolayers, we solved the BSE equation on top of G0W0 calculation. The imaginary part of the dielectric function and corresponding oscillator strength of the optical transitions of HG monolayer is depicted in Figure \ref{opt-transition}. In this case, we have two excitonic states with a very strong oscillator strength. The starting two peaks appears at 1.28, and 1.52 eV in the optical spectrum of HG monolayer (see Figure \ref{opt-transition}) mainly originate from the optical transitions at the high symmetry $\Gamma$ point in the BZ. The second peak splitted with 0.24 eV of optical transitions is also originate at the same high symmetry $\Gamma$ point. It was reported that the band gap is 1.36 eV using self-energy corrections within the GW method by Moreno et. al.\cite{moreno2018bottom}.  According to that the excitonic binding energy (EBE) of HG monolayer is found to be 80 meV which is calculated by; EBE=$E_{g}^{GW}$-$E_{g}^{optical}$.

\begin{figure}[t]
	    \centering
	    \includegraphics[width=1.0\linewidth]{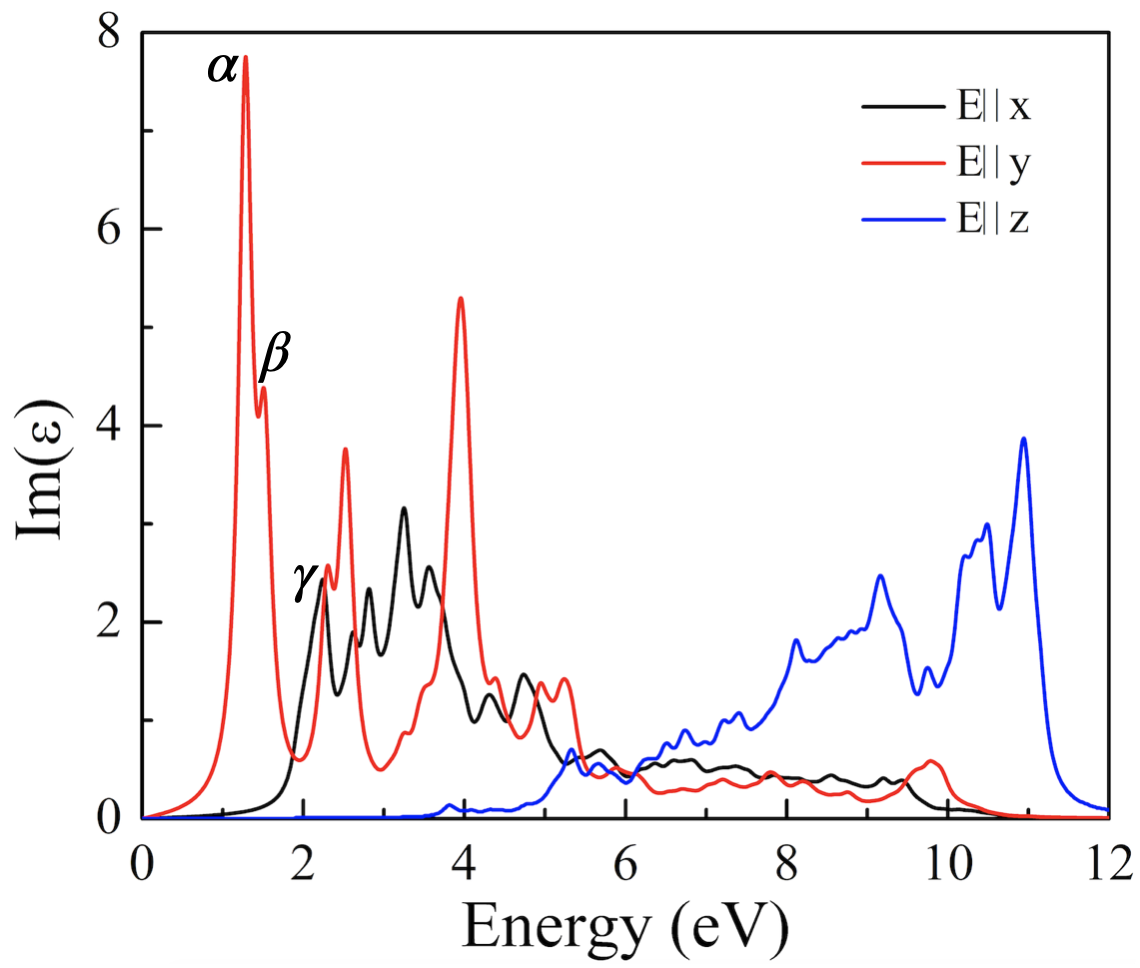}
	    \caption{The imaginary part of the dielectric function E$\parallel$X, E$\parallel$y and E$\parallel$z corresponding to the holey graphene monolayer along the x, y and z directions, respectively. The black, red and blue lines represent the imaginary components along x, y and z directions, respectively. The z component is the multiple of 100 on y-axis data.}
	    \label{anisotropic-img}
    \end{figure}

It is clear that the optical anisotropy among E$\parallel$X, E$\parallel$y and E$\parallel$z  largely increases by the inclusion of local field effects, as shown in Figure \ref{anisotropic-img}.  The anisotropic results shows along E$\parallel$X and E$\parallel$y means in-plane, the allowed optical transitions observed at 1.28 eV ($\alpha$), 1.52 eV ($\beta$) and 2.24 eV ($\gamma$). According to optical selection rules, only $\pi$ $\rightarrow$ $\pi^{*}$ and $\sigma$ $\rightarrow$ $\sigma^{*}$ transitions are allowed if the light polarized parallel to the planar directions. It means that the mentioned peaks occurs from $\pi$ $\rightarrow$ $\pi^{*}$ transition. Moreover, along the z direction i.e., E$\parallel$z, out-of-plane, the allowed optical transitions appears at 3.8 eV and 5.32 eV. These transition peaks are mainly originate from $\pi$ $\rightarrow$ $\sigma^{*}$, or $\sigma$ $\rightarrow$ $\pi^{*}$ transitions which is allowed for perpendicular polarization direction i.e. E$\parallel$z, out-of-plane. Such types of spectra are useful as they provide valuable information on optical transition probability corresponding to certain light wavelength.

\begin{figure}[t]
	    \centering
	    \includegraphics[width=1.0\linewidth]{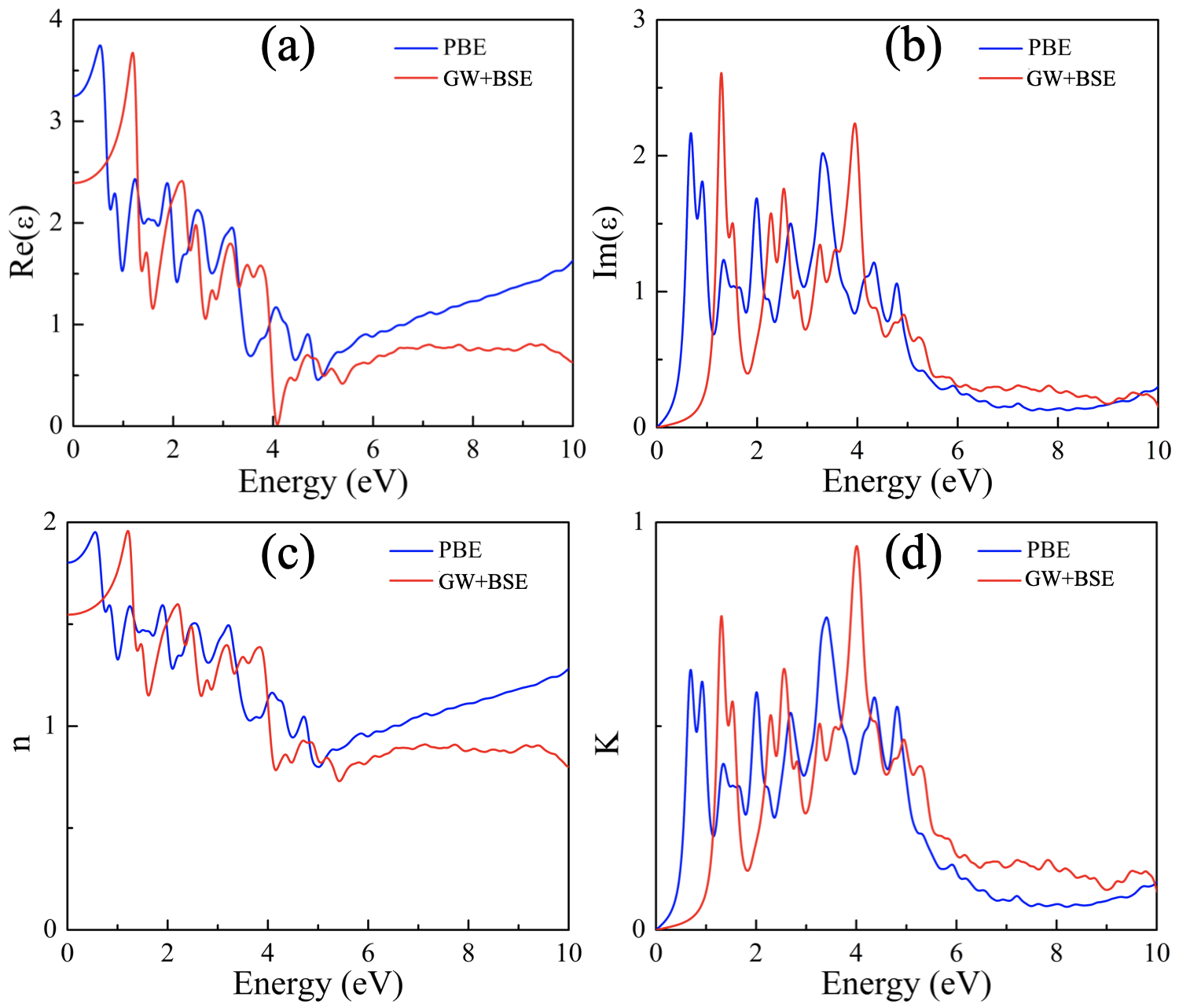}
	    \caption{Optical properties of holey graphene: (a) real part of the dielectric function, (b) imaginary part of the dielectric function, (c) refractive index and (d) extinction coefficient with the comparison of PBE functional and GW+BSE method.}
	    \label{opt-1}
    \end{figure}

Furthermore, the real part of dielectric function shows the  material polarizability. The static values of the real part of the complex dielectric function at $\omega$=0 is depicted in Figure \ref{opt-1}(a) are 2.40 (with GW+BSE) and 3.25 (with PBE). From real part of dielectric function, (-)ve values appears at $\approx$ 4 eV. It shows the metallic character of HG monolayer in the ultraviolet (UV) part of the electromagnetic spectrum. Also, we have compared the imaginary part of dielectric function with two methods PBE and GW plus BSE, in which GW plus BSE peaks shifted towards higher photon energy range and compare to PBE as shown in Figure \ref{opt-1}(b). 

Figure \ref{opt-1}(c) illustrates the computed refractive index of HG monolayer using PBE, and GW plus BSE. The value of the static refraction index n(0) in the HG sheet for is 1.54, and 1.81 using GW plus BSE and PBE, respectively. It is well known that the GW plus BSE method gives very accurate optical properties. According to that refractive index of HG monolayer is 1.54. It means that HG monolayer is fully transparent material because its refractive index is equal to glass refractive index.  From the Figure \ref{opt-1}(c), we can observe that at energies about 4 eV using both method, the refractive index is minimum and at that values the absorption is maximum. The refractive index, n increases with photon energy in the infrared (IR) region, while decreases monotonically in the visible and UV region and then it steadily decreases. The maximum refractive index is found to be 1.95 at 1.28 ev for HG monolayer with GW plus BSE method. This refractive indices could enable HG monolayer to be used as a optical cavity layer\cite{matsko2009practical}. Additionally, we have calculated the extinction coefficient, K as shown in Figure \ref{opt-1}(d). The values of extinction coefficient, K of HG monolayer rapidly decrease with increasing photon energy in the UV region and become oscillatory type of behaviour after 5 eV. The maximum values of the extinction coefficient found at 4 eV. It means that at this energies the photons will be absorbed very fast (i.e. penetration depth will be the shortest at particular photon energy)\cite{singh2016antimonene}.
     
\begin{figure}[t]
	    \centering
	    \includegraphics[width=1.0\linewidth]{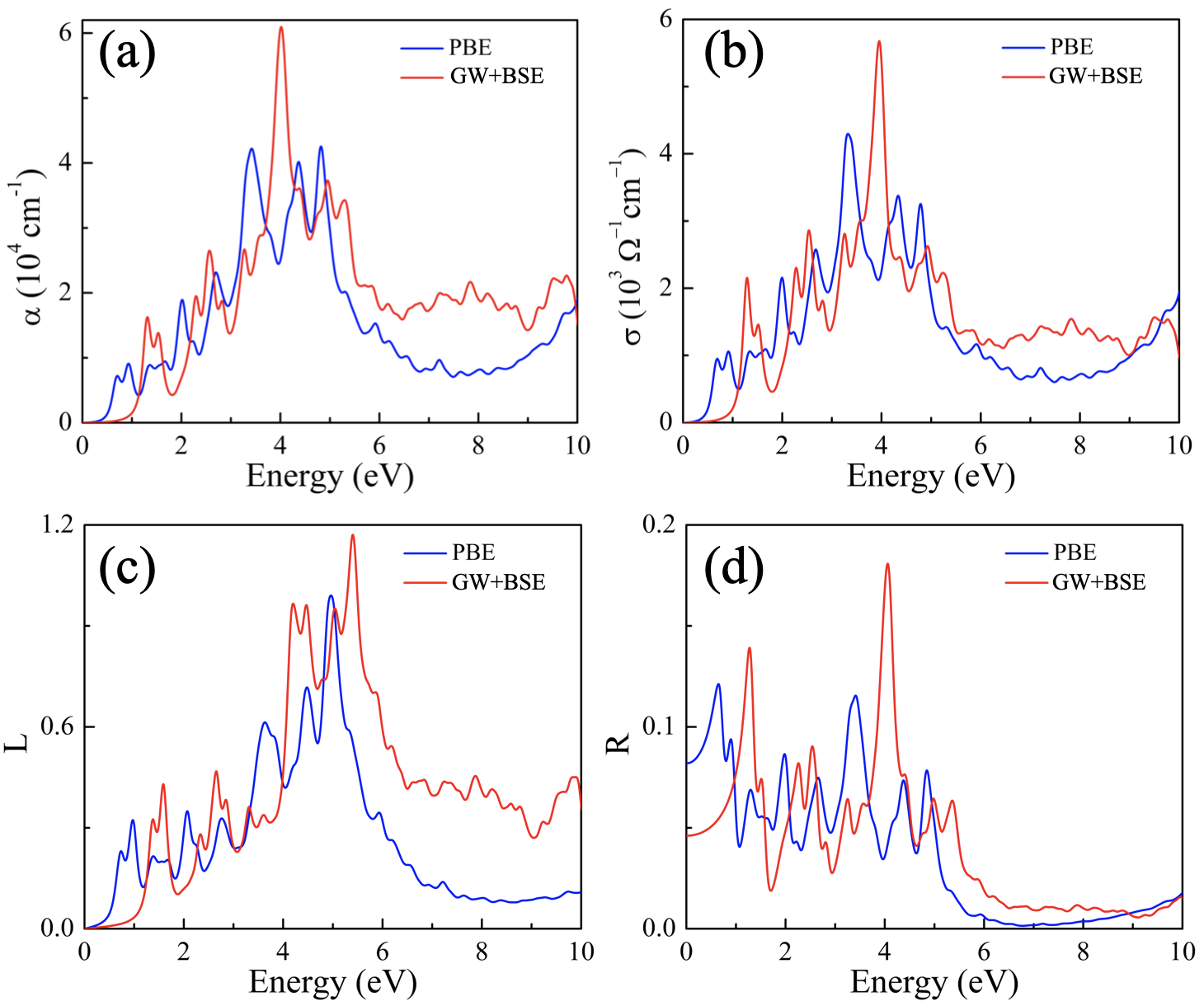}
	    \caption{Optical properties of holey graphene: (a) absorption spectra, (b) optical conductivity, (c) electron energy loss function and (d) reflectively with the comparison of PBE functional and GW+BSE method.} 
	    	    \label{opt-2}
    \end{figure}

Now, we further discuss about the optical absorption spectra shown in Figure \ref{opt-2}(a). The absorption spectra of HG monolayer computed by PBE and GW plus BSE method. When the PBE+RPA were included, then the entire absorption shifted to the lower photon energy i.e. it shows the red shift, while with e-h interactions  included (i.e. GW+BSE), the whole spectrum is shifted towards higher photon energy. Henceforth, the blue shift of optical absorption occurs due to the excitonic effects. We can see that the absorption starts from low photon energy of 0.65 eV with PBE plus RPA and slightly shifted higher energy at 1.28 eV with GW+BSE methods. The absorption coefficient of HG monolayer reaches to its maximum value of 3.5 eV and 4 eV with PBE+RPA and GW+BSE, respectively. We found the maximum values of optical conductivity at same photon energy (like maximum optical absorption) as shown in Figure \ref{opt-2}(b). After the 4 eV of photon energy, the optical absorption and conductivity is drastically decreases with increasing photon energy and shows oscillatory behaviour. Also, the electron energy loss spectra, EELS (L) is very useful in realize plasma resonance phenomena as distinct from normal interband transitions. Figure \ref{opt-2}(c) shows the L of HG monolayer with two method. Furthermore,there is a large sharp peak in photon energy at 1.52 eV with GW+BSE method, and the same peak were found at very low photon energy with PBE+RPA method. This peak at lower photon energy comes from the Frenkel exciton\cite{ellis2008charge, baldini2017strongly} which is related to $\pi$ plasmon in the L. The second strong and sharp peak appears at 5.5 eV which is strongly related to $\pi+\sigma$ electron plasmon\cite{shahrokhi2016quasi}. The reflectivity of HG monolayer is presented in Figure \ref{opt-2}(d). It is notice that the reflectivity in HG monolayer shows more oscillatory nature up to 5.5 eV, after that it rapidly decreases with increasing photon energy. The maximum reflectivity is found to be 18 $\%$ around 4 eV. The reflectivity values not exceed 20 $\%$ in whole region that why it can be used in anti-reflective coating. While the reflectivity was found to be 40 $\%$ in monolayer graphene by Qiu and co-workers \cite{qiu2018optical}, which is two times larges than HG monolayer.

\subsection{Thermoelectric properties}
\label{thermo}
\begin{figure}[t]
	    \centering
	    \includegraphics[width=1.0\linewidth]{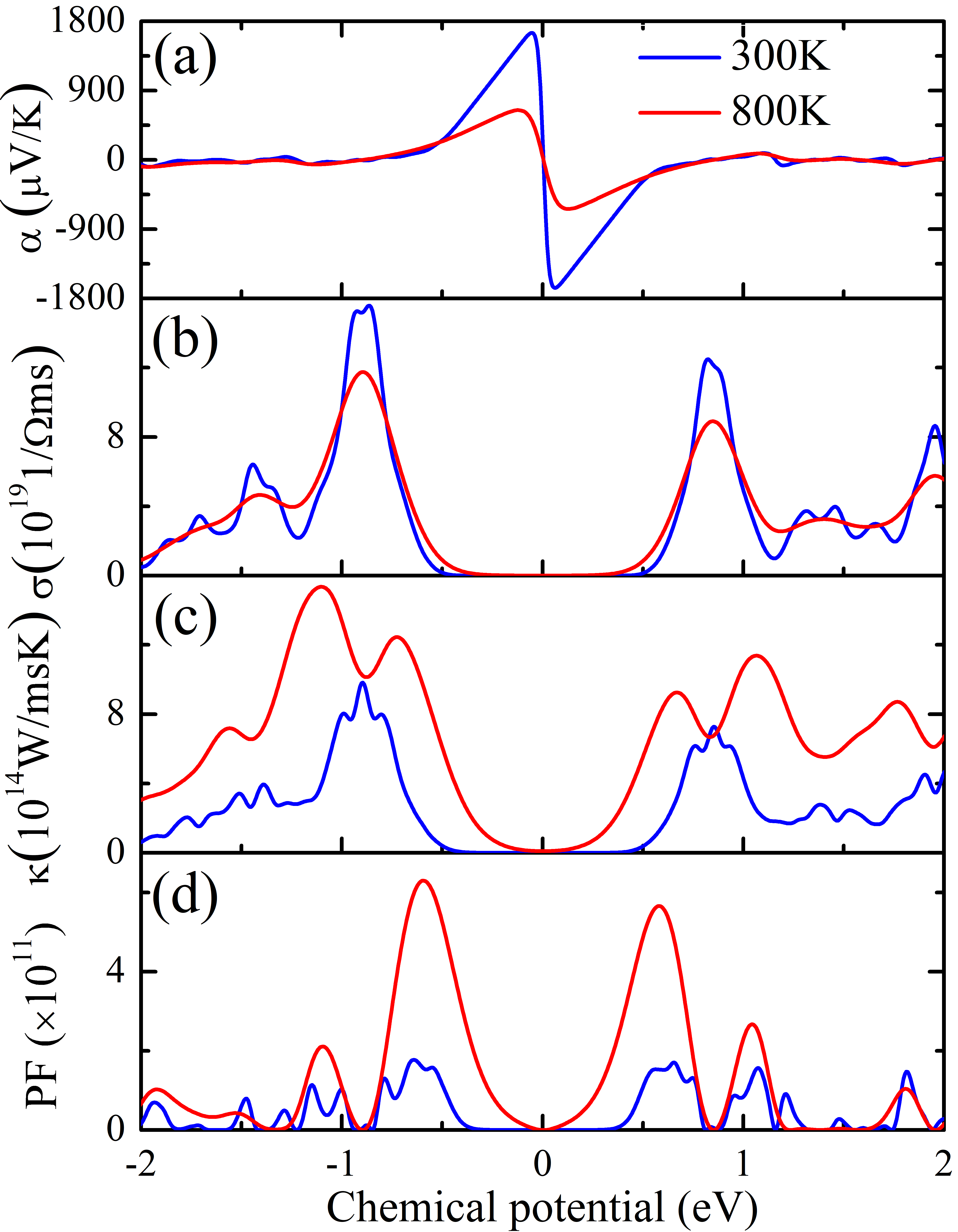}
	    \caption{Thermoelectric component as a function of chemical potential in the range of -2 eV to +2 eV. (a) Seebeck coefficient, (b) electrical conductivity, (c) electronic thermal conductivity and power factor of HG monolayer.} 
	    	    \label{thermo-avg}
    \end{figure}

In this section, we will discuss the thermoelectric properties such as Seebeck coefficient $\alpha$, electrical conductivity $\sigma$, electronic thermal conductivity $\kappa_{e}$, thermopower, PF and electronic figure of merit, ZT as a function of chemical potential for HG monolayer. The thermopower, $\alpha$ of HG monolayer as a function of chemical potential, $\mu$ is presented in Figure \ref{thermo-avg}(a) at two different temperatures  300K and 800K. The values of $\mu$ in (-)ve and (+)ve side show p-type and n-type doping of the system, respectively. The each term of thermoelectric properties got enhanced characteristics in p-type and n-type doping in HG monolayer. From Figure \ref{thermo-avg}(a), we can observe that the sharp values of thermopower around the Fermi level (i.e. p-type and n-type doping side) and suppressed at higher temperature, which indicates an optimal carrier concentration is favorable for enhancing the thermoelectric performance. The maximum values of thermopower at room temperature is found to be 1662.59 $\mu$V/K in n-type doping and 645.67 $\mu$V/K in p-type doping at 800K. Because the magnitude of thermopower values is higher in that region. The thermopower graph is almost similar values on both side p-type and n-type doping due to the symmetric nature of valence and conduction bands (see Figure \ref{F2}).

The electrical conductivity is depicted in Figure \ref{thermo-avg}(b), shows the values of $\sigma$/$\tau$ as a function of chemical potential with increasing behaviour near to the Fermi level. The increasing thermopower correlates with the gradual decrease in electrical conductivity in p-type and n-type doping, which clearly reflects the electronic band structure. The electrical conductivity is also sensitive similar to the thermopower, $\sigma$/$\tau$ exibit relatively lower value at 800 K temperature with enhanced value near the Fermi level (see Figure \ref{thermo-avg}(b)). Overall the electrocal conductivity is larger in p-type doping than n-type around the Fermi level.
In addition to this the variation of electronic thermal conductivity with chemical potential is presented in Figure \ref{thermo-avg}(c). The thermal conductivity is more sensitive with temperature. The electronic thermal conductivity shows similar trend to that electrical conductivity. Similar to electric conductivity the values of $\kappa_{e}$ is relatively higher in p-type doping in HG monolayer.

    \begin{figure}[t]
	    \centering
	    \includegraphics[width=1.0\linewidth]{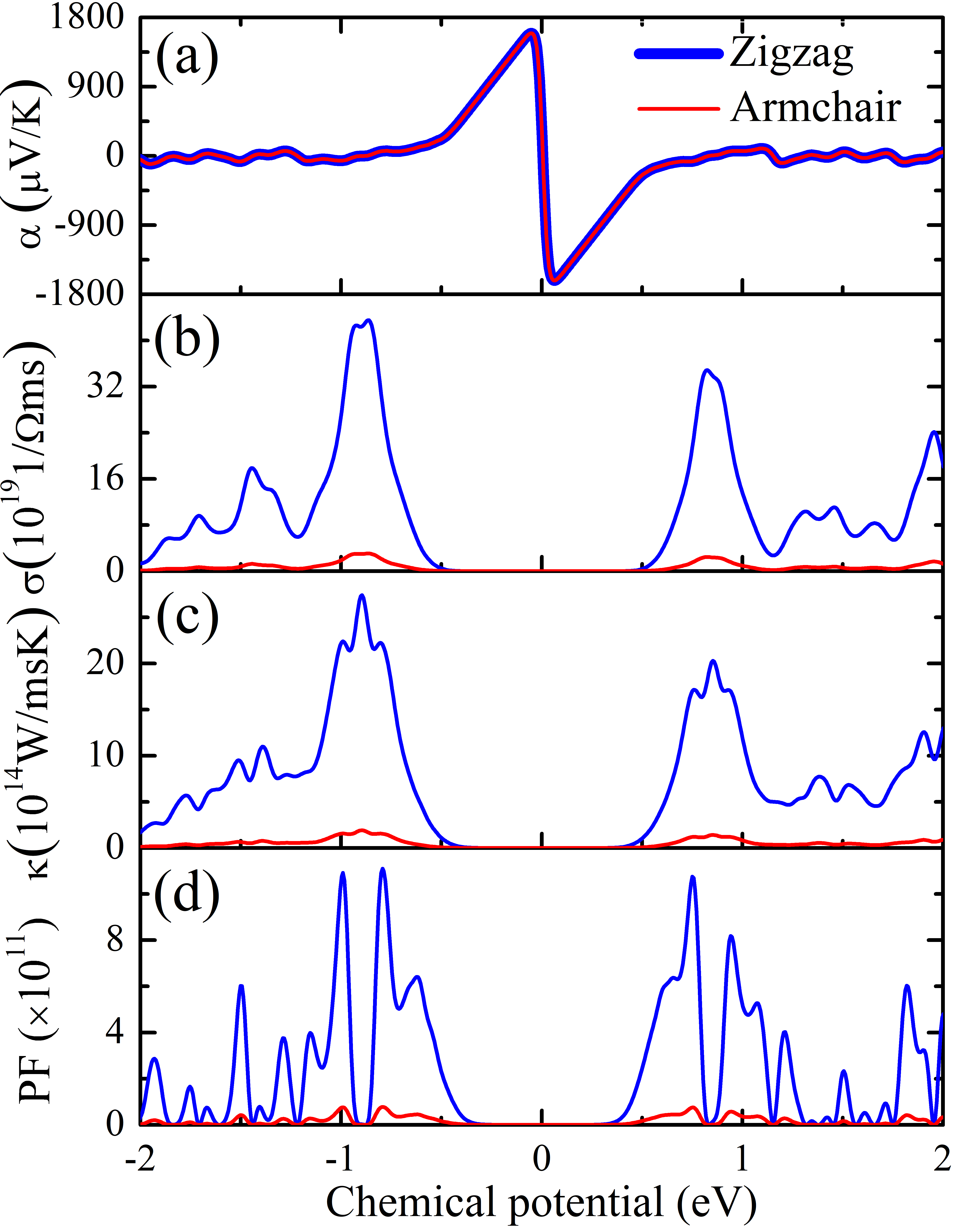}
	    \caption{The directional dependence of thermoelectric component i.e. zigzag and armchair direction as a function of chemical potential in the range of -2 eV to +2 eV. (a) Seebeck coefficient, (b) electrical conductivity, (c) electronic thermal conductivity and power factor of HG monolayer.} 
	    \label{thermo-anisotropic}
    \end{figure}

The power factor, PF ($\sigma \alpha^{2}$) of HG monolayer is shown in Figure \ref{thermo-avg}(d). The computed the values of $\sigma \alpha^{2}$ at two different temperature is depicted as a function of chemical potential. The maximum PF mainly increases with band gap decreasing which could be associated to the increased electrical conductivity\cite{bi2018thermoelectric}. Similar to previous results the PF also shows larger values in p-type doping as compare to n-type doping  at higher temperature and it is increases with doping level increasing. Additionally, the sharp electronic density of states near the Fermi level or flat band lines in the electronic band structure near the Fermi level enhanced the thermopwer $\alpha$. Further the high DOS near Fermi energy provides large electrical conductivity.


Furthermore, we have calculated the directional dependence of thermopower $\alpha$, electrical conductivity $\sigma$, electronic thermal conductivity $\kappa_{e}$, PF as a function of chemical potential as shown in Figure \ref{thermo-anisotropic}. The thermopower is depend on the sharp DOS or flat band lines near the Fermi level. That why the thermopower values in the zigzag and armchair direction is same (see Figure \ref{thermo-anisotropic}(a)). While the electrical conductivity, electronic thermal conductivity and power factor are significantly affected for directions, which are presented in Figure \ref{thermo-anisotropic}.

    \begin{figure}[t]
	    \centering
	    \includegraphics[width=1.0\linewidth]{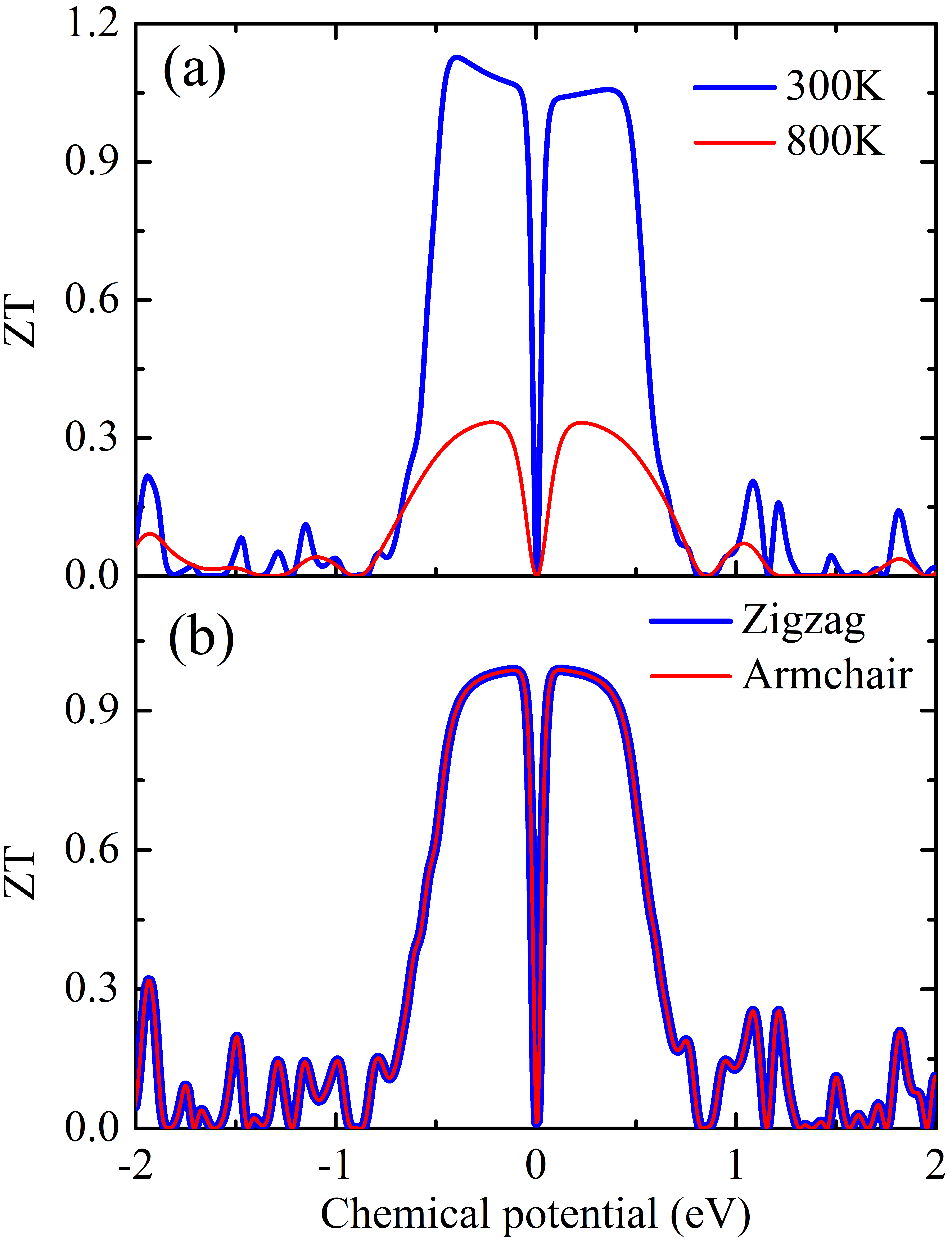}
	    \caption{Electronic figure of merit, ZT (a) average and (b) zigzag and armchair direction of HG monolayer.} 
	    	    \label{ZT}
    \end{figure}

From all the transport quantities, we have calculated the electronic figure of merit, ZT as a function of chemical potential, at fixed temperatures of 300 K and 800 K for HG monolayer. At room temperature, ZT reach maximum in p-type doping levels and corresponding value of 1.13 is observed at chemical potential of -0.40 eV. Whereas at higher temperature 800 K, ZT value of 0.33 at chemical potential of $\pm$0.22 eV for both p- and n-type doping. Is it good or band in comparison of these materials. The values of ZT was 1.02 for boron monochalcogenide\cite{mishra2020two}, $\approx$ 0.38 in CP monolayer\cite{singh2018single}, $\approx$ 0.75 for arsenene monolayer\cite{sharma2017arsenene} and $\approx$ 0.78 for antimonene monolayer\cite{sharma2017arsenene} and 0.08 for single layer of graphene\cite{reshak2014thermoelectric} in the previous investigation. Furthermore, the figure of merit is not affected in zigzag and armchair directions and remains same. It means that the electronic figure of merit exhibits isotropic behaviour. One can see that the p-type doping level enhanced the thermpower as compared to n-type doping levels. Thus one can conclude that p-type doping in HG monolayer should be potential thermoelectric materials at room temperature.

\section{Conclusion}
\label{conclusion}
In summary, we systematically investigated the electronic, optical and thermoelectric properties of recently synthesis novel 2D planar HG monolayer by using first-principles calculations. To find the accurate optical properties, we have used GW plus BSE methods. In HG structure, strong excitonic effects play a crucial role in optical properties, with a significant small binding energy assigned to bound excitons. We found that the hybrid functional HSE06 band gap is 0.95 eV, closer to the experimental value than the results from other reported works. Also, the optical gap of 1.28 eV was obtained and an excitonic binding energy of 0.08 eV. The broad absorptions in the visible light region are found for novel 2D HG monolayer. Interestingly the reflectivity does not exceed more than 20 \% which lower than graphene. This suggests that this material can be used as a anti-reflective coating. Moreover, we found the superior thermopower values of 1662.59 $\mu$V/K and corresponding electronic figure of merit of 1.13 in HG monolayer larger than most of the 2D materials and graphene monolayer. We suggest monolayered HG possesses extraordinary electronic, optical and thermoelectric properties. Our results are timely and claims that HG is suitable for opto and thermo electronic device applications.

  
\begin{acknowledgements} 
    DS and RA thanks  Olle Engkvists stiftelse, Carl Tryggers Stiftelse for Vetenskaplig Forskning (CTS) and Swedish Research Council (VR) for financial support. SNIC and HPC2N are acknowledged for providing the computing facilities.
\end{acknowledgements} 
  

\bibliographystyle{apsrev4-1}

%
\end{document}